\documentclass[11pt]{article}
\usepackage{epsfig}
\usepackage{amsmath}
\usepackage{amssymb}

\begin{document}

\title{Study of Pure Annihilation  Decays $B_{d,s} \to D^{0} \overline D^{0}$ }

\author{Ying Li\footnote{e-mail: liying@ytu.edu.cn} \ and Juan Hua
 \\
{\it  Physics Department, Yantai University, Yantai 264005, China}}
 \maketitle

\begin{abstract}
With heavy quark limit and hierarchy approximation $\lambda_{QCD}\ll
m_D\ll m_B$, we analyze the $B\to D^0\overline D^0$ and $B_s\to
D^0\overline D^0$ decays, which occur purely via annihilation type
diagrams. As a roughly estimation, we calculate their branching
ratios and CP asymmetries in Perturbative QCD approach. The
branching ratio of $B\to D^0\overline D^0$ is about
$3.8\times10^{-5}$ that is just below the latest experimental upper
limit. The branching ratio of $B_s\to D^0\overline D^0$ is about
$6.8\times10^{-4}$, which could be measured in LHC-b. From the
calculation, it could be found that this branching ratio is not
sensitive to the weak phase angle $\gamma$. In these two decay
modes, there exist CP asymmetries because of interference between
weak and strong interaction. However, these asymmetries are too
small to be measured easily.
\end{abstract}

\section{Introduction}\label{sc:intro}

In the Standard Model (SM), CP-violation (CPV) arises from a complex
phase in the Cabibbo-Kobayashi- Maskawa (CKM) quark mixing matrix,
and the angles of unitary triangle are defined as \cite{Yao:2006px}:
\begin{eqnarray}
\beta=\arg \Bigl[-\frac{V_{cb}^*V_{cd}}{V_{tb}^*V_{td}}\Bigl],~~~~
\alpha=\arg \Bigl[-\frac{V_{tb}^*V_{td}}{V_{ub}^*V_{ud}}\Bigl],~~~~
\gamma=\arg
\Bigl[-\frac{V_{ub}^*V_{ud}}{V_{cb}^*V_{cd}}\Bigl].\label{ckmangle}
\end{eqnarray}
In order to test SM and search for new physics, many measurements of
CP-violation observables can be used to constrain these above
angles. It is well known that we   measure $\beta$ precisely using
the golden   decay mode $B\to J/\psi K_s$; the angle $\alpha$ can be
determined with decay $B\to\pi\pi$ and $\gamma$ could be measured
precisely in Large Hadron Collider (LHC) with  decay mode $B_s\to
D_sK$.

\begin{figure}[thb]
\begin{center}
\includegraphics[scale=0.8]{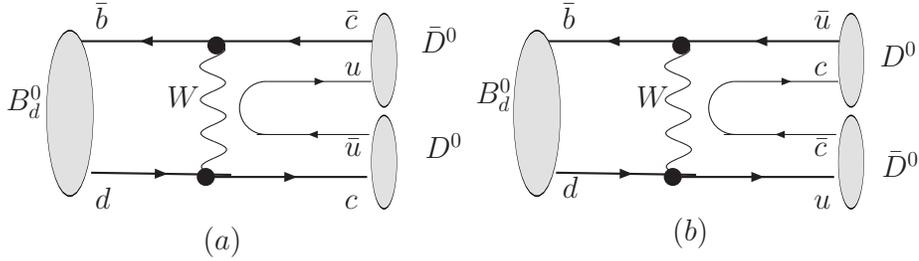}
\caption{The quark level Feynman diagrams for $B_d \to D^{0} \bar
D^{0}$ process } \label{fig0}
\end{center}
\end{figure}

Besides the above channels mentioned, many other channels are used
to cross check the measurements. Among these decays, $B\to DD$ decay
is considered to test the $\beta$ measurement. For $B\to DD$ decay,
the analysis based on $SU(3)$ symmetry \cite{Savage:1989ub},
iso-spin symmetry \cite{Xing:1999yx} and factorization approach
\cite{Xing:1998ca} have been done in last several years. However,
the the calculation of decay $B^0 \to D^0\overline D^0$ has
difficulties. It is a pure-annihilation diagram decay, also named
W-exchange diagram decay, which is power suppressed in factorization
language. The quark diagrams of this decay are shown in Figure
\ref{fig0}. Theoretically, QCD factorization approach (QCDF)
\cite{Beneke:1999br} and soft collinear effective theory
(SCET)\cite{Bauer:2001yt} can not deal decays with two heavy charmed
mesons  effectively. In Ref.\cite{Keum:2003js,Lu:2003xc},
perturbative QCD (PQCD) has been exploited to $B$ meson decays with
one charm meson in the final states and the results agree with
experimental data well. Specially, the pure  annihilation type B
decays with charmed meson  were studied  in Ref.\cite{Lu:2003xc}.

In the standard model picture, the $W$ boson exchange causes
$\bar{b}d \to \bar{c}c $, and the $\bar{u}u$ quarks are produced
from a gluon. This gluon attaches to any one of the quarks
participating in the $W$ boson exchange. In decay $B \to D^{0}
\overline D^{0}$, the momentum of the final state $D$ meson is
$\frac{1}{2} m_B (1-2 r^2)$, with $r=m_D/m_B$. If we consider heavy
quark limit and hierarchy approximation $\lambda_{QCD}\ll m_D\ll
m_B$, the $D$ meson momentum is nearly $m_B/2$. According to the
distribution amplitude used in Ref.\cite{Keum:2003js}, the light
quark in $D$ meson carrying nearly $40\%$ of the $D$ meson momentum.
So, this light quark is still a collinear quark with 1
$\mathrm{GeV}$ energy, like that in $B\to DM$
\cite{Keum:2003js,Lu:2003xc}, $B\to K(\pi)\pi$ \cite{kls,luy}
decays. The gluon could be viewed as a hard gluon approximatively,
so we can treat the process perturbatively where the four-quark
operator exchanges a hard gluon with $u \bar u$ quark pair. Of
course, we are able to calculate the diagrams if charm quark and up
quark exchange.  As a roughly estimation, we give the branching
ratio and CP-violation of $B_{d,s} \to D^{0} \overline D^{0}$.

In this article, the analytic formulas for the decay amplitudes will
be shown in the next section. In section \ref{sc:result}, we give
the numerical results and summarize this article in section
\ref{sc:summary}.

\section{Analytic formulas}\label{sc:analy}

For simplicity, we set $B$ meson at rest in our calculation. In
light-cone coordinates, the momentum of $B$, $D^0$ and$\overline
D^0$
 are:
\begin{eqnarray}
P_B=\frac{M_B}{\sqrt{2}}(1,1,\vec{0});
P_2=\frac{M_B}{\sqrt{2}}(1-r^2,r^2,\vec{0});
P_3=\frac{M_B}{\sqrt{2}}(r^2,1-r^2,\vec{0}).
\end{eqnarray}
we define the light (anti-)quark momenta in $B$, $D^0$ and
$\overline D^0$ mesons as $k_1$, $k_2$, and $k_3$ as:
 \begin{equation} k_1 =
(x_1P_1^+,0,{\bf k}_{1T}),\ \  k_2 = (x_2 P_2^+,0,{\bf k}_{2T}),\ \
 k_3 = (0, x_3 P_3^-,{\bf k}_{3T}). \label{eq:momentun2}
\end{equation}

In PQCD, we factorize the decay amplitude into soft($\Phi$),
hard($H$), and harder ($C$) dynamics characterized by different
scales, \cite{kls,luy}
\begin{multline}
\mathcal{A}\sim
 \int\!\! d x_1 d x_2 d x_3 b_1 d b_1 b_2 d b_2 b_3
d b_3  \mathrm{Tr} \Bigl[ C(t) \Phi_B(x_1,b_1) \Phi_{D}(x_2,b_2)
\Phi_D(x_3, b_3) H(x_i, b_i,t) S_t(x_i)\, e^{-S(t)} \Bigr].
\label{eq:convolution2}
\end{multline}
In above equation, $b_i$ is the conjugate space coordinate of the
transverse momentum ${\bf k}_{iT}$, and $t$ is the largest energy
scale. $C$ is Wilson coefficient, and $\Phi$ is the wave function.
The last term, $e^{-S(t)}$, contains two kinds of contributions. One
 is due to the resummation of  the large double logarithms from renormalization of
ultra-violet divergence $\ln tb$, the other is from resummation of
double logarithm $\ln^2 b$ from the overlap of collinear and soft
gluon corrections, which is called Sudakov form factor. The hard
part $H$ can be calculated perturbatively, and it is channel
dependent. More explanation of above formula and review about PQCD
can be found in many reference, such as \cite{kls,luy,Ali:2007ff}.

As a heavy meson, the $B$ meson wave function is not well defined,
neither is $D$ meson. In heavy quark limit, we take them as:
\begin{equation}
 \Phi_{B}(x,b) = \frac{i}{\sqrt{6}}
\left[ \not \! P  + M_B  \right] \gamma_5\phi_B(x,b),
 \end{equation}
 \begin{equation}
 \Phi_{D}(x,b) = \frac{i}{\sqrt{6}}\gamma_5
\left[ \not \! P  +  M_D \right] \phi_D(x,b).
\end{equation}
The Lorentz structure of two mesons are different because the $B$
meson is initials state and $D$ meson is final state.

The effective Hamiltonian  $\bar b\to \bar q(q=d,s)$ is given by
\cite{Buchalla:1996vs}:
\begin{multline}
 \mathcal{H}_\mathrm{eff} =
 \frac{G_F}{\sqrt{2}}\Bigl\{ V_{cq}V_{cb}^* \Bigl[
C_1(\mu) O_1^c(\mu) + C_2(\mu) O_2^c(\mu) \Bigl]+ V_{uq}V_{ub}^*
\Bigl[ C_1(\mu) O_1^u(\mu) + C_2(\mu) O_2^u(\mu) \Bigl]\\
-V_{tb}^*V_{tq}\sum_{i=3}^{10}C_i(\mu) O_i(\mu)\Bigl\},\label{hami}
\end{multline}
 where $C_{i}(\mu)(i=1,\cdots,10)$ are Wilson coefficients at the
  renormalization scale $\mu$ and the four quark operators $O_{i}(i=1,\cdots,10)$ are
\begin{equation}\begin{array}{ll}
  O_1^c = (\bar{b}_ic_j)_{V-A}(\bar{c}_jq_i)_{V-A},  &
  O_2^c = (\bar{b}_ic_i)_{V-A} (\bar{c}_jq_j)_{V-A},  \\
  O_1^u = (\bar{b}_iu_j)_{V-A}(\bar{u}_jq_i)_{V-A},  &
  O_2^u = (\bar{b}_iu_i)_{V-A} (\bar{u}_jq_j)_{V-A},  \\
  O_3 = (\bar{b}_iq_i)_{V-A}\sum_{q} (\bar{q}_jq_j)_{V-A},  &
  O_4 = (\bar{b}_iq_j)_{V-A}\sum_{q} (\bar{q}_jq_i)_{V-A}, \\
  O_5 = (\bar{b}_iq_i)_{V-A}\sum_{q} (\bar{q}_jq_j)_{V+A},  &
  O_6 = (\bar{b}_iq_j)_{V-A} \sum_{q} (\bar{q}_jq_i)_{V+A}, \\
  O_7 = \frac{3}{2}(\bar{b}_iq_i)_{V-A} \sum_{q}
   e_q(\bar{q}_jq_j)_{V+A},   &
   O_8 = \frac{3}{2}(\bar{b}_iq_j)_{V-A}\sum_{q} e_q
  (\bar{q}_jq_i)_{V+A}, \\
  O_9 = \frac{3}{2}(\bar{b}_iq_i)_{V-A}\sum_{q}
  e_q(\bar{q}_jq_j)_{V-A}, &
   O_{10} = \frac{3}{2}(\bar{b}_iq_j)_{V-A}\sum_{q}
 e_q(\bar{q}_jq_i)_{V-A}. \label{eq:effectiv}
 \end{array}
\end{equation}
Here $i$ and $j$ are $SU(3)$ color indices; in $O_{3,...,10}$ the
sum over $q$ runs over the quark fields that are active at the scale
$\mu=O(m_{b})$, i.e., $q\in \{u,d,s,c,b\}$. For Wilson coefficients,
we will also use the leading logarithm summation for QCD
corrections, although the next-to -leading order calculation already
exists \cite{Buchalla:1996vs}. This is the consistent way to cancel
the explicit $\mu$ dependence in the theoretical formulae.

\begin{figure}[thb]
\begin{center}
\includegraphics[scale=0.6]{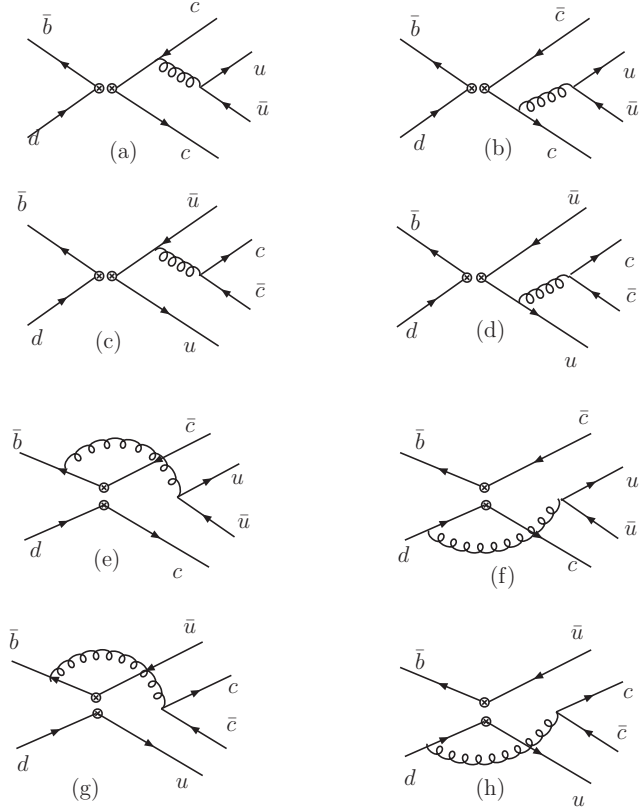}
\caption{The leading order Feynman diagrams for $B_d \to D^{0}
\overline D^{0}$ process in PQCD approach} \label{fig1}
\end{center}
\end{figure}
According to the effective Hamiltonian in eq.(\ref{hami},
\ref{eq:effectiv}), the lowest order diagrams of $B \to D^0\overline
D^0$ are drawn in Fig. \ref{fig1}. We first calculate the usual
factorizable diagrams (a), (b), (c) and (d). For the $(V-A)(V-A)$
operators, their contributions of (a) and (c) are always canceled by
diagrams (b) and (d) respectively because of current conservation.
For the $(V-A)(V+A)$ operators, these diagrams can not give
contribution, either. That's to say, factorizable diagrams have no
contribution. For non-factorizable diagrams (e), (f), (g) and (h),
we find the hard part for $(V-A)(V-A)$ operators are same to
$(V-A)(V+A)$ operators. We group the contribution of diagrams (e)
and (f), denoted by $M_a$, as follows:
\begin{multline}
M_{a}[C_i]  =  \frac{64 \pi C_FM_B^2}{\sqrt{2N_C}}  \int_0^1 \!\!
dx_1 dx_2 dx_3
 \int_0^\infty \!\! b_1 db_1\, b_2 db_2\
\phi_B(x_1,b_1) \phi_{D}(x_2,b_2)\phi_{D}(x_3,b_2) \\
\times \Bigl\{ \Bigl[x_1+x_2+(2x_3-x_2)r^2\Bigl]
C_i(t_{a}^1)E(t_{a}^1) h_a^{(1)}(x_1,
x_2,x_3,b_1,b_2) \\
+\Bigl[-x_3+(2x_1-2x_2+x_3)r^2\Bigl] C_i(t_{a}^2)E(t_{a}^2)
h_a^{(2)}(x_1, x_2,x_3,b_1,b_2) \Bigr\}, \label{eq:Ma}
\end{multline}
where $C_F = 4/3$ is the group factor of $\mathrm{SU}(3)_c$ gauge
group, and $C_i$ is Wilson coefficient. The function  $E_m$ is
defined as
 \begin{equation}
  E(t) = \alpha_s(t)\, e^{-S_B(t)-S_D(t)-S_D(t)},
 \end{equation}
and $S_B$, $S_D$ result from Sudakov factor and single logarithms
due to the renormalization of ultra-violet divergence. The functions
$h_a$ is the Fourier transformation of virtual quark and gluon
propagators. It is defined by
\begin{align}
 &h^{(j)}_a(x_1,x_2,x_3,b_1,b_2) = \nonumber \\
 & \biggl\{ \frac{\pi i}{2}
 \mathrm{H}_0^{(1)}(M_B\sqrt{x_2x_3(1-2r^2)}\, b_1)
  \mathrm{J}_0(M_B\sqrt{x_2x_3(1-2r^2)}\, b_2) \theta(b_1-b_2)
 \nonumber \\
 & \qquad\qquad\qquad\qquad + (b_1 \leftrightarrow b_2) \biggr\}
  \times\left(
 \begin{matrix}
  \mathrm{K}_0(M_B F_{a(j)} b_1), & \text{for}\quad F^2_{a(j)}>0 \\
  \frac{\pi i}{2} \mathrm{H}_0^{(1)}(M_B\sqrt{|F^2_{a(j)}|}\ b_1), &
  \text{for}\quad F^2_{a(j)}<0
 \end{matrix}\right),
 \label{eq:propagator1}
 \end{align}
with:
\begin{eqnarray}
  F^2_{a(1)} &=&-x_1-x_2-x_3+x_1x_3+x_2x_3+(x_2+x_3-x_1x_3-2x_2x_3)r^2;\\
  F^2_{a(2)}&=&x_2x_3-x_1x_3+(x_1x_3-2x_2x_3)r^2.
\end{eqnarray}
In above equation, $\mathrm{H}_0^{(1)}(z) = \mathrm{J}_0(z) + i\,
 \mathrm{Y}_0(z)$. In order to reduce the large logarithmic radiative
corrections, the hard scale $t$ in the amplitudes is selected as the
largest energy scale in  the hard part:
 \begin{eqnarray}
  t_{a}^j &=& \mathrm{max}(M_B \sqrt{|F^2_{a(j)}|},
 M_B \sqrt{(1-2r^2)x_2x_3 }, 1/b_1,1/b_2).
 \end{eqnarray}
Analogically, we can get the $M_{b}$, which comes from the
contribution of diagrams (g) and (h):
\begin{multline}
M_{b}[C_i]=  \frac{64 \pi C_FM_B^2}{\sqrt{2N_C}}  \int_0^1 \!\! dx_1
dx_2 dx_3
 \int_0^\infty \!\! b_1 db_1\, b_2 db_2\
\phi_B(x_1,b_1) \phi_{D}(x_2,b_2)\phi_{D}(x_3,b_2) \\
\times \Bigl\{ \Bigl[1-x_3+(2+2x_1-2x_2+x_3)r^2\Bigl]C_i(t_{b}^1)
E(t_{b}^1) h_b^{(1)}(x_1,
x_2,x_3,b_1,b_2) \\
+\Bigl[x_1+x_2-1+(-2-x_2+2x_3)r^2\Bigl]C_i(t_{b}^2) E(t_{b}^2)
h_b^{(2)}(x_1, x_2,x_3,b_1,b_2) \Bigr\}, \label{eq:Mb}
\end{multline}
and the functions are defined as:
  \begin{align}
 &h^{(j)}_b(x_1,x_2,x_3,b_1,b_2) = \nonumber \\
 & \biggl\{ \frac{\pi i}{2}
 \mathrm{H}_0^{(1)}(M_B\sqrt{1-x_2-x_3+x_2x_3+(x_2+x_3-2x_2x_3)r^2}\,
 b_1) \nonumber\\
&~~~~~~~~
\times\mathrm{J}_0(M_B\sqrt{1-x_2-x_3+x_2x_3+(x_2+x_3-2x_2x_3)r^2}\,
b_2) \theta(b_1-b_2)
 \nonumber \\
 & \qquad\qquad\qquad\qquad + (b_1 \leftrightarrow b_2) \biggr\}
  \times\left(
 \begin{matrix}
  \mathrm{K}_0(M_B F_{b(j)} b_1), & \text{for}\quad F^2_{b(j)}>0 \\
  \frac{\pi i}{2} \mathrm{H}_0^{(1)}(M_B\sqrt{|F^2_{b(j)}|}\ b_1), &
  \text{for}\quad F^2_{b(j)}<0
 \end{matrix}\right);
 \label{eq:propagator2}
 \end{align}
 \begin{eqnarray}
  F^2_{b(1)}&=&-1-x_1x_3+x_2x_3+(x_1x_3-2x_2x_3)r^2,\\
  F^2_{b(2)}&=&1-x_1-x_2-x_3+x_1x_3+x_2x_3+(x_2+x_3-x_1x_3-2x_2x_3)r^2,
 \end{eqnarray}
 \begin{eqnarray}
  t_{b}^j &=& \mathrm{max}(M_B \sqrt{|F^2_{b(j)}|},
 M_B \sqrt{1-x_2-x_3+x_2x_3+(x_2+x_3-2x_2x_3)r^2 }, 1/b_1,1/b_2).
  \end{eqnarray}
So, the decay amplitude of decay $B_d \to D^{0} \overline D^{0}$ can
be read as:
 \begin{eqnarray}
\mathcal{A}_1&=&V_{cb}^*V_{cd}M_{a}[C_2]-V_{tb}^*V_{td}M_{a}[C_5+C_7]+V_{ub}^*V_{ud}M_{b}[C_2]
-V_{tb}^*V_{td}M_{b}[C_5+C_7]\nonumber\\
           &=&V_{cb}^*V_{cd}T_1
           -V_{tb}^*V_{td}P_1\nonumber\\
           &=&V_{tb}^*V_{td}P_1(1+z_1e^{i(\beta+\delta_1)}),\label{bdd1}
\end{eqnarray}
where $\beta$ is weak phase angle defined in Eq.(\ref{ckmangle}),
and $\delta_1$ is the strong phase, which plays an important role in
studying CP-violation. In above calculation, we denote that
\begin{eqnarray}
T_1&=&M_{a}[C_2]-M_{b}[C_2], \nonumber \\
P_1&=&M_{a}[C_5+C_7]+M_{b}[C_5+C_7]+M_{b}[C_2],
\end{eqnarray}
and
\begin{eqnarray}
z_1=\left|\frac{V_{cb}^*V_{cd}}{V_{tb}^*V_{td}}
\right|\left|\frac{T_1}{P_1}\right|,
\end{eqnarray}
which describes the ratio between tree diagram and penguin diagram.
The corresponding charge conjugate decay is
\begin{eqnarray}\label{bdd2}
\overline{\mathcal{A}_1}=V_{tb}V_{td}^*P_1(1+z_1e^{i(-\beta+\delta_1)}).
\end{eqnarray}
Therefore, the averaged  decay width $\Gamma$ for $B^0 \to  D^0
\overline D^0$ decay is then given by
\begin{equation}
 \Gamma(B^0 \to  D^0
\overline D^0) = \frac{G_F^2 M_B^3}{128\pi}
 (1-2r^2)|V_{tb}^*V_{td}P_1|^2
\bigl|1+z_1^2+2z_1\cos \beta \cos\delta_1|. \label{eq:neut_width1}
\end{equation}
>From this equation, we know that the averaged branching ratio is a
function of CKM angle $\beta$, if $z_1\neq 0$.  Derived from
Eq.(\ref{bdd1}) and Eq.(\ref{bdd2}), the direct CP-violation can be
formulated as:
\begin{eqnarray}\label{cp1}
A_{CP}^{dir}(B\to D^{0} \overline D^{0})=\frac{|A_{B_d\to
D^0\overline D^0}|^2-|A_{\overline B_d\to \overline
D^0D^0}|^2}{|A_{B_d\to D^0\overline D^0}|^2+|A_{\overline B_d\to
\overline
D^0D^0}|^2}=\frac{-2z_1\sin\beta\sin\delta_1}{1+z_1^2+2z_1\cos \beta
\cos\delta_1}.
 \end{eqnarray}

 For $ B^0_s \to  D^0 \overline D^0$ and its conjugate decay, we write the decay amplitudes  and
 rearrange them as:
 \begin{eqnarray}
\mathcal{A}_2&=&V_{cb}^*V_{cs}M_{a}[C_2]-V_{tb}^*V_{ts}M_{a}[C_5+C_7]+V_{ub}^*V_{us}M_{b}[C_2]
-V_{tb}^*V_{ts}M_{b}[C_5+C_7]\nonumber\\
            &=&V_{ub}^*V_{us}M_{b}[C_2]-V_{tb}^*V_{ts}\Bigl\{M_{a}[C_5+C_7]
            +M_{b}[C_5+C_7]-\frac{V_{cb}^*V_{cs}}{V_{tb}^*V_{ts}}M_{a}[C_2]\Bigl\}\nonumber\\
            &=&V_{ub}^*V_{us}T_2-V_{tb}^*V_{ts}P_2\nonumber\\
            &=&V_{ub}^*V_{us}T_2\Bigl[1+z_2e^{i(-\gamma+\delta_2)}\Bigl],\\
\overline{\mathcal{A}_2}&=&V_{ub}V_{us}^*T_2\Bigl[1+z_2e^{i(\gamma+\delta_2)}\Bigl],\label{bdd3}
\end{eqnarray}
where $T_2$, $P_2$ and $z_2$ are defined as:
\begin{eqnarray}
T_2&=&M_{b}[C_2],\nonumber\\
 P_2&=&M_{a}[C_5+C_7]
            +M_{b}[C_5+C_7]-\frac{V_{cb}^*V_{cd}}{V_{tb}^*V_{ts}}M_{a}[C_2],\nonumber\\
z_2&=&\left|\frac{V_{tb}^*V_{ts}}{V_{ub}^*V_{us}}
\right|\left|\frac{T_2}{P_2}\right|.
\end{eqnarray}
So, the averaged decay width and direct CP violation can be
formulated as:
\begin{eqnarray}\label{eq:neut_width2}
 \Gamma(B_s \to  D^0\overline D^0) &=& \frac{G_F^2 M_B^3}{128\pi} (1-2r^2) \bigl|V_{ub}V_{us}^*T_2\bigr|^2
 (1+z_2^2+2z_2\cos\delta_2\cos\gamma),\\
A_{CP}^{dir}(B_s \to D^0\overline D^0)&=&\frac{|A_{B_s\to
D^0\overline D^0}|^2-|A_{\overline B_s\to \overline
D^0D^0}|^2}{|A_{B_s\to D^0\overline D^0}|^2+|A_{\overline B_s\to
\overline
D^0D^0}|^2}=\frac{2z_2\sin\gamma\sin\delta_2}{1+z_2^2+2z_2\cos
\gamma \cos\delta_2}.
\end{eqnarray}
In our calculation, we set $m_c\approx m_D$, just because $m_D -m_c
\approx \Lambda_{QCD}$ and $\frac{\Lambda_{QCD}}{m_B} \rightarrow 0$
in the heavy quark limit.

\section{Numerical Results}\label{sc:result}

For $B$ meson, the distribution amplitude is well determined by
charmless $B$ decays \cite{kls,luy}, which is chosen as
\begin{equation}
\phi_B(x,b) = N_B x^2(1-x)^2 \exp \left[ -\frac{M_B^2\ x^2}{2
\omega_b^2} -\frac{1}{2} (\omega_b b)^2 \right],\label{waveb}
\end{equation}
with parameters  $\omega_b=0.4\mbox{ GeV}$, and $N_B=91.745\mbox{
GeV}$ which is the normalization constant using $f_B=190 \mbox{
MeV}$. For $B_s$ meson, we use the same wave function according to
SU(3) symmetry, where $\omega_b=0.4\mbox{ GeV}$, $N_{B_s}=119.4
\mbox{ GeV}$ and $f_{B_s}=230 \mbox{ MeV}$.

Since the $c$ quark is much heavier than the $u$ quark, the $c$
quark shares more momentum, and this function should be asymmetric
with respect to $x = 1/2$. The asymmetry is parameterized by $a_D$.
Similar to the $b$-dependence on the wave function of $B$ meson, for
controlling the size of charmed mesons, we also introduce the
intrinsic $b$-dependence on those of charmed mesons. Hence, we use
the wave function    of $D$ meson as \cite{Chen:2003px}
\begin{equation}
\phi_{D}(x,b) = \frac{3}{\sqrt{2 N_c}} f_{D} x(1-x)\Bigl[ 1 +
a_{D} (1 -2x)\Bigl]\exp \left[-\frac{1}{2} (\omega_D b)^2
\right].\label{waved}
\end{equation}
We use $a_{D}=0.7$ and $\omega_{D}=0.4$ in above function. Other
parameters, such as meson mass, decay constants, the CKM matrix
elements and the lifetime of $B$ meson are list
\cite{Yao:2006px,Follana:2007uv}:
\begin{gather}
M_B = 5.28 \mbox{ GeV},\  M_{B_S} = 5.36 \mbox{ GeV},\  M_{D} = 1.87
\mbox{ GeV},\ f_{D} = 210
\mbox{ MeV}, \nonumber \\
|V_{ud}|=0.974, \  |V_{ub}|= 4.3 \times 10^{-3},\   |V_{cd}|=0.23,
\   |V_{cb}|= 41.6\times 10^{-3} \nonumber\\
|V_{td}|= 7.4\times 10^{-3}, \  |V_{tb}|=1.0,\
|V_{us}|=0.226, \  |V_{cs}|=0.957,\nonumber \\
|V_{ts}|=41.6\times 10^{-3}, \quad\tau_{B_d^0}=1.54\times 10^{-12}\mbox{ s,}\quad
 \tau_{B_s^0}=1.46\times 10^{-12}\mbox{ s}.\label{parameterZ}
\end{gather}

\begin{table}
  \caption{Amplitudes ($10^{-3}$ GeV) of $B_d\to D^0\overline D^0$ and
   $B_s\to D^0\overline D^0$.}\label{tb:amplitudes}
 \begin{center}
  \begin{tabular}[b]{r|cc}
   \hline
   \hline
            &$B_d\to D^0\overline D^0$ &$B_s\to D^0\overline D^0$ \\
   \hline
   \hline
   $T(e)+T(f)$ & $68+17 \, i$ &$66 +27 \, i$\\
   $P(e)+P(f)$ & $ 0.80+ 0.23\, i$ &$0.77 +3.68\, i$\\
   $T(g)+T(h)$ & $ 9.81- 2.99\, i$ &$14.0-0.6\, i$\\
   $P(g)+P(h)$ & $ 0.08- 0.02\, i$ &$-0.01 +0.01\, i$ \\
   \hline
   \hline
  \end{tabular}
 \end{center}
\end{table}

\begin{figure}[thb]
\begin{center}
\includegraphics[scale=1.0]{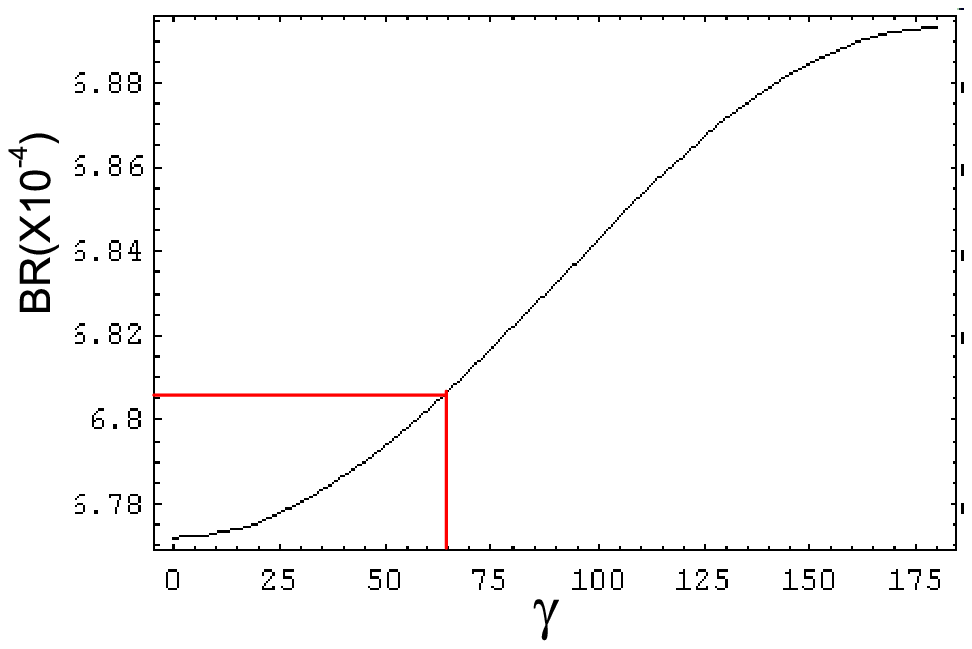}
\caption{The branching ratio of $B_s\to D^0\overline D^0$ changes
with CKM angle $\gamma$.} \label{fig3}
\end{center}
\end{figure}

\begin{figure}[thb]
\begin{center}
\includegraphics[scale=0.6]{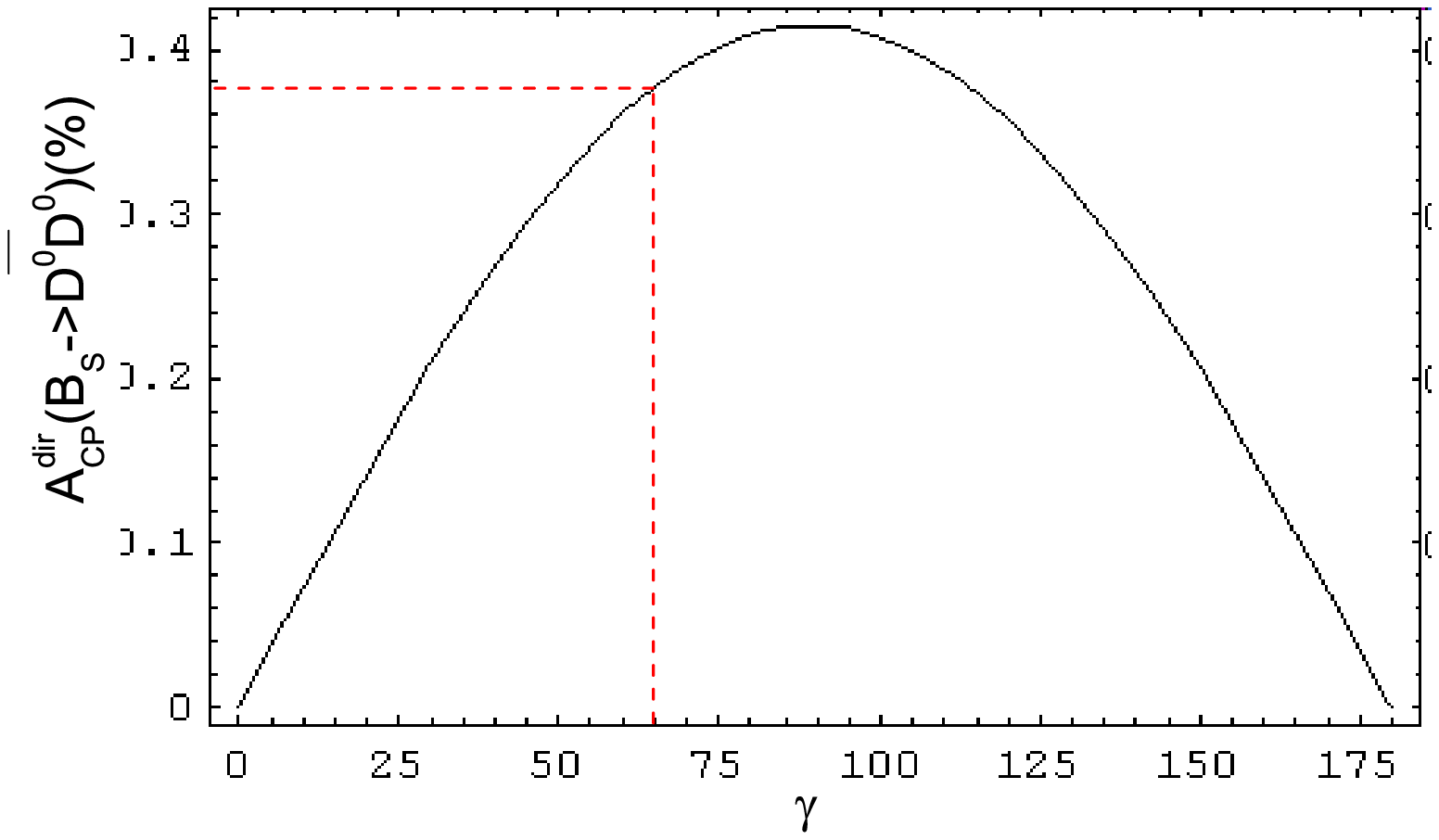}
\caption{The direct CP-violation of $B_s\to D^0\overline D^0$
changes with CKM angle $\gamma$.} \label{fig4}
\end{center}
\end{figure}

With these parameters fixed, we calculate the decay amplitudes  of
the $B^0\to D^0\overline D^0$ and $B_s\to D^0\overline D^0$ decays
in Table~\ref{tb:amplitudes}. From the table, we notice that the
main contribution comes from the tree diagram (e) and (f). And our
predictions for the branching ratio of each mode corresponding to
$\beta=23^\circ$ and $\gamma=63^\circ$ are listed,
\begin{eqnarray}
  BR(B_d\to D^0\overline D^0) &=& 2.3\times 10^{-5}; \nonumber\\
  BR(B_s\to D^0\overline D^0) &=& 6.8\times 10^{-4}.
\end{eqnarray}
In Fig. \ref{fig3}, we plot the branching ratio of $B_s\to
D^0\overline D^0$ with different $\gamma$. In this figure, we find
the branching ratio is not sensitive to  CKM angle $\gamma$. For the
experimental side, there are only upper limits given at $90\%$
confidence level for decay $B_d\to D^0\overline D^0$,
\begin{eqnarray}
  BR(B_d\to D^0\overline D^0) &< &6.0\times 10^{-5}; ~~~~~~BarBar\cite{Aubert:2006ia}\nonumber\\
  BR(B_d\to D^0\overline D^0) &< &4.2\times 10^{-5}. ~~~~~~Belle\cite{:2007sk}
\end{eqnarray}
Obviously, our result is consistent with the data. For $B_d\to
D^0\overline D^0$ decay mode, $z_1$ is about 6.5, and the strong
phase $\delta_1$ is $34^\circ$,  so $A_{CP}^{dir}$ is about $-6\%$
with the definition in Eq.(\ref{cp1}). As decay mode $B_s\to
D^0\overline D^0$ is concerned, $z_2$ is about $205$ and
$\delta_2=155^\circ$, and the relation between direct CP violation
and $\gamma$ is shown in Fig.\ref{fig4}. From the figure, we  read
the CP asymmetry is about $0.4\%$, which is rather tiny. It is
necessary to state that the $z_1$ and $z_2$ are not the true ratio
between tree contribution and penguin, because mathematical technics
are used in Eq. (\ref{bdd1}) and (\ref{bdd3}).

In addition to the perturbative annihilation contributions, there is
also a hadronic picture for the $B_d\to D^0\overline D^0$,  named
soft final states interaction\cite{Cheng:2004ru}. The $B$ meson
decays into $D^+$ and $D^-$, the secondary particles then exchanging
a $\rho$ meson, then scatter into $D^0\overline D^0$ through final
state interaction afterwards. For $B_s$ decay, the $B_s$ meson
decays into $D_s^+$ and $D^+$ then scatters into $D^0\overline D^0$
by exchanging a Kaon. But this picture cannot be calculated
accurately because of lack of many effective vertexes, and we will
ignore this contribution here, though it may be important
\cite{Cheng:2004ru}.

\begin{table}[ht]
\caption{The sensitivity of the decay branching ratios and CP
asymmetries to change  of ¥ø$\omega_b$, $\omega_D$ and
$a_D$}\label{table} {\small
\begin{center}
\begin{tabular}[t]{rcccc}
\hline
                  &$BR(B_d\to D^0\overline D^0)$ & $BR(B_s\to D^0\overline D^0)$
                  &$A_{CP}^{dir}(B_d\to D^0\overline D^0)$
                  &$A_{CP}^{dir}(B_s\to D^0\overline D^0)$ \\
                  &$(\times 10^{-5})$ & $(\times 10^{-4})$&$(\%)$&$(\%)$ \\
\hline
$\omega_b(B\setminus B_s)$          & & &  &      \\
$0.35\setminus0.45$       &4.3&7.8& -7.2&0.4\\
$0.40\setminus0.50$       &3.8&6.8& -5.3&0.4\\
$0.45\setminus0.55$       &3.2&5.9& -5.8&0.4\\
\hline
$\omega_D$       &  &   &   &       \\
$0.35$           &5.0&9.7&-4.2&0.3\\
$0.40$           &3.8&6.8&-5.3&0.4\\
$0.45$           &2.2&4.2&-7.8&0.5\\
\hline
$a_D$            &   &   &   &      \\
$0.6$            &3.2&5.9&-6.9&0.4\\
$0.7$            &3.8&6.8&-5.3&0.4\\
$0.8$            &4.3&7.8&-6.1&0.4\\
 \hline
\end{tabular}
\end{center}}
\end{table}

There are many uncertainties in our calculation such as higher order
corrections, the parameters listed in Eq.(\ref{parameterZ}) and the
distribution amplitudes of heavy mesons. We will not discuss
uncertainty taken by high order correction as we only roughly
estimate the branching ratios and CP asymmetries, though high order
corrections have been done for some special channels
\cite{Li:2005kt,Li:2006jv} and showed $15-20\%$ uncertainty. The
parameters in Eq.(\ref{parameterZ}), fixed by experiments, are
proportional to the amplitudes, so we will not analyze this kind
uncertainty either. In our calculation, we find that the results are
sensitive to the distribution amplitudes, especially to that of $D$
meson. Since the heavy $D$ wave function is less constrained, we set
$a_{D}\in (0.6-0.8)\mbox{ GeV}$ and $\omega_{D}\in(0.35-0.45)\mbox{
GeV}$ to exploit the uncertainty. Table~\ref{table} shows the
sensitivity of the branching ratios to change of $\omega_b$,
$\omega_D$ and $a_{D}$. It is found that uncertainty of the
predictions on PQCD is mainly due to $\omega_D$, which describes the
behavior in end-point region of $D$ meson, however it is very hard
to be determined. Considering the experimental upper limit, our
results favor large $\omega_b$, large $\omega_D$ and small $a_D$.

At last, we give the prediction of branching ratios with err bar as
follows:
\begin{eqnarray}
  BR(B_d\to D^0\overline D^0) &=&(3.8^{+0.5+1.2+0.5}_{-0.6-1.6-0.6})\times 10^{-5}
  \left(\frac{f_B\cdot f_D\cdot f_D}{190\mathrm{ GeV}\cdot 210\mathrm{ GeV}\cdot 210\mathrm{ GeV}}\right)^2;  \nonumber\\
  BR(B_s\to D^0\overline D^0) &=&(6.8^{+1.0+2.9+1.0}_{-0.9-2.6-0.9})\times 10^{-4}
  \left(\frac{f_{B_s}\cdot f_D\cdot f_D}{230\mathrm{ GeV}\cdot 210\mathrm{ GeV}\cdot 210\mathrm{ GeV}}\right)^2.
\end{eqnarray}
We believe that the $B_d\to D^0\overline D^0$ will be measured soon
because this ratio is just below the upper limit, and $B_d\to
D^0\overline D^0$ will be measured in LHC-b in next year as a
channel to cross check the $\gamma$ measurements.

\section{Summary}\label{sc:summary}

With heavy quark limit and hierarchy approximation $\lambda_{QCD}\ll
m_D\ll m_B$, we analyze the $B\to D^0\overline D^0$ and $B_s\to
D^0\overline D^0$ decays, which occur purely via annihilation type
diagrams. As a roughly estimation, we calculate the branching ratios
and CP asymmetries in PQCD approach. The branching ratios are still
sizable. The branching ratio of $B\to D^0\overline D^0$ is about
$3.8\times10^{-5}$, which is just below the experimental upper
limited result\cite{Aubert:2006ia, :2007sk}, and we think that it
will be measured in near future. For  $B_s\to D^0\overline D^0$, the
branching ratio is about $6.8\times10^{-4}$, which could be measured
in LHC-b. From the calculation, it is found that this branching
ratio is not sensitive to angle $\gamma$. In these two decays, there
exist CP asymmetries because of interference between weak and strong
interaction, though they are very small.

\section*{Acknowledgements}

Y.Li thanks  Institute of High Energy Physics for their hospitality
during  visit where part of this work was done. This work is partly
supported  by Foundation of Yantai University under Grant
No.WL07B19. We would like to acknowledge C.D. L\"u and W. Wang for
valuable discussions.

\end{document}